\documentclass{article}
\usepackage{epsfig}

\date{\today}

\usepackage{amsmath}

\usepackage{graphicx}

\newcommand{\insertplot}[5]{\begin{figure}
 \hfill\hbox to 0.05in{\vbox to #5in{\vfill
 \inputplot{#1}{#4}{#5}}\hfill}
 \hfill\vspace{-.1in}
 \caption{#2}\label{#3}
 \end{figure}}

 \newcommand{\inputplot}[3]{
 \special{ps: plotfile #1}
}
\newcounter{fig}   

\newcounter{fixy}

\begin{document}

 \newenvironment{fixy}[1]{\setcounter{figure}{#1}}
{\addtocounter{fixy}{1}}
\renewcommand{\thefixy}{\arabic{fixy}}
\renewcommand{\thefigure}{\thefixy\alph{figure}}
\setcounter{fixy}{1}

\title{Non Abelian Chern-Simons--Higgs solutions in $2+1$ dimensions}
\author{{\large Francisco Navarro-L\'erida,}$^{\ddagger}$
{\large Eugen Radu}$^{\diamond}$
and {\large D. H. Tchrakian}$^{\star}$ \\ 
$^{\ddagger}${\small Dept.de F\'isica At\'omica, Molecular y Nuclear, Ciencias F\'isicas,}\\
{\small Universidad Complutense de Madrid, E-28040 Madrid, Spain}\\  
$^{\diamond}${\small Institut f\"ur Physik, Universit\"at Oldenburg, Postfach 2503, D-26111 Oldenburg, Germany}\\ 
$^{\star}${\small School of Theoretical Physics -- DIAS, 10 Burlington
Road, Dublin 4, Ireland }}

\date{}
\newcommand{\dd}{\mbox{d}}
\newcommand{\tr}{\mbox{tr}}
\newcommand{\la}{\lambda}
\newcommand{\ka}{\kappa}
\newcommand{\f}{\phi}
\newcommand{\vf}{\varphi}
\newcommand{\F}{\Phi}
\newcommand{\al}{\alpha}
\newcommand{\ga}{\gamma}
\newcommand{\de}{\delta}
\newcommand{\si}{\sigma}
\newcommand{\Si}{\Sigma}
\newcommand{\bomega}{\mbox{\boldmath $\omega$}}
\newcommand{\bsi}{\mbox{\boldmath $\sigma$}}
\newcommand{\bchi}{\mbox{\boldmath $\chi$}}
\newcommand{\bal}{\mbox{\boldmath $\alpha$}}
\newcommand{\bpsi}{\mbox{\boldmath $\psi$}}
\newcommand{\brho}{\mbox{\boldmath $\varrho$}}
\newcommand{\beps}{\mbox{\boldmath $\varepsilon$}}
\newcommand{\bxi}{\mbox{\boldmath $\xi$}}
\newcommand{\bbeta}{\mbox{\boldmath $\beta$}}
\newcommand{\ee}{\end{equation}}
\newcommand{\eea}{\end{eqnarray}}
\newcommand{\be}{\begin{equation}}
\newcommand{\bea}{\begin{eqnarray}}

\newcommand{\ii}{\mbox{i}}
\newcommand{\e}{\mbox{e}}
\newcommand{\pa}{\partial}
\newcommand{\Om}{\Omega}
\newcommand{\vep}{\varepsilon}
\newcommand{\bfph}{{\bf \phi}}
\newcommand{\lm}{\lambda}
\def\theequation{\arabic{equation}}
\renewcommand{\thefootnote}{\fnsymbol{footnote}}
\newcommand{\re}[1]{(\ref{#1})}
\newcommand{\R}{{\rm I \hspace{-0.52ex} R}}
\newcommand{\N}{{\sf N\hspace*{-1.0ex}\rule{0.15ex}%
{1.3ex}\hspace*{1.0ex}}}
\newcommand{\Q}{{\sf Q\hspace*{-1.1ex}\rule{0.15ex}%
{1.5ex}\hspace*{1.1ex}}}
\newcommand{\C}{{\sf C\hspace*{-0.9ex}\rule{0.15ex}%
{1.3ex}\hspace*{0.9ex}}}
\newcommand{\eins}{1\hspace{-0.56ex}{\rm I}}
\renewcommand{\thefootnote}{\arabic{footnote}}

\maketitle


\bigskip

\begin{abstract}
Non Abelian vortices of a $SU(2)$ Chern-Simons--Higgs theory in $2+1$ dimensions are constructed numerically. They
represent natural counterparts of the U(1) solutions considered by Hong, Kim and Pac, and, by Jackiw and Weinberg.
The Abelian
embeddings are identified, for all values of the Higgs selfinteraction strength $\nu$, resulting in both attractive and
repulsive phases. 
A detailed analysis of the properties of the solutions reveals the existence of a number of unexpected features.
For a certain range of the parameter $\nu$, it is shown that the non Abelian vortices have lower
energy than their topologically stable Abelian counterparts, resulting in an effective energy lower bound on the
$SU(2)$ configurations. The angular momentum of these vortices is analysed and it is found that unlike the Abelian ones,
whose angular momentum and energy are unrelated, there is a nontrivial mass--spin relation of the non Abelian
vortices. 
\end{abstract}
\medskip
\medskip

\section{Introduction}
Selfdual Abelian Chern-Simons--Higgs vortices were constructed by Hong, Kim and Pac~\cite{Hong:1990yh}, and, by
Jackiw and Weinberg~\cite{Jackiw:1990aw} describing anyonic solitons in $2+1$ dimensions. These excited considerable
interest because of their relevance to high $T_c$ superconductivity and resulted in the development of a large body of
literature, including also  the case  when the Maxwell term was present. For a complete review we refer to
\cite{Dunne:1998qy} (see also the recent work \cite{Horvathy:2008hd}).

The study of these Abelian Chern-Simons--Higgs (CS-H) vortices was motivated by the discovery in ~\cite{Deser:1982vy}, of
topologically massive (non Abelian) Yang--Mills (YM) theories augmented by a Chern--Simons (CS) term. Thus it is that
even before the discovery of the former~\cite{Hong:1990yh,Jackiw:1990aw}, their non Abelian versions were considered.
Early work employing a non Abelian Yang-Mills--CS-H (YM-CS-H) model featuring a pair of Higgs fields in the adjoint
representations of $SU(2)$ was carried out in
\cite{deVega:1986eu,Kumar:1986yz}, and in the adjoint representations of 
$SU(N)$ in \cite{deVega:1986hm}. Non Abelian CS-H vortices with Higgs
field in the fundamental representation of $SU(2)$ were also discussed in \cite{Lee:1990ep}.
Recently solutions in a supersymmetric ${\cal N}=2$ non Abelian CS-Higgs (CS-H)
theory were considered in \cite{Aldrovandi:2007nb}, and  more recently, vortex solutions
to a $U(N)$ YM-CS-H system with adjoint representation Higgs, describing the bosonic sector of a ${\cal N}=2$
supersymmetric model, were given in \cite{Collie:2008za} where the moduli space approach of \cite{Manton:1981mp} was
employed to study the vortex dynamics at low energy.

However,  it appears that in the existing literature on this subject no numerical constructions of non Abelian CS-H
vortices are presented. Given the absence of analytic solutions in closed form, this means that to date no concrete
constructions are given.
In the present work, we consider the simplest possible $SU(2)$ extension of the $2+1$ dimensional Abelian CS-H model of
\cite{Hong:1990yh,Jackiw:1990aw}, with adjoint representation Higgs. 
The Abelian embedding of these $SU(2)$ vortices in the selfdual
limit are none other than the selfdual vortices of \cite{Hong:1990yh,Jackiw:1990aw}.

While the physical interest of such vortices may
be in condensed matter physics, here we shall put the main emphasis on the numerical study of these solutions, 
and not on their physical application.
Therefore, our approach is just the opposite and therefore 
complementary to the one generally adopted in the existing literature.

Our study focuses on the full non Abelian solutions with special attention given to the relation of these with their
Abelian counterparts. The reason is that there is no Bogomol'nyi type topological lower bound for the simple non Abelian
system we consider, in contrast to the Abelian subsystem for which there exists the (magnetic) vortex number, which is
the only topological charge, resulting in the topological stability of the Abelian vortices.
It is then pertinent to examine whether the
energies of the corresponding non Abelian vortices, namely those characterised by the same values of the physical
coupling constants in the model, are larger or smaller than those of the Abelian embeddings. This would have a bearing
on the question of the stability of the non Abelian vortices. To this end, we have displayed the energies of the
vorticity$-n$ solutions versus the Higgs selfinteraction parameter. This is an important question since these non
Abelian solutions do not saturate a Bogomol'nyi bound, unlike for example in the presence of massive fundamental
scalars~\cite{Collie:2008za}.

Another interesting question addressed is that of the value of the angular momentum of the non Abelian vortices,
which is the only global quantity exclusively characterising these,
relative to the known values for their Abelian counterparts~\cite{Hong:1990yh,Jackiw:1990aw}.

Finally, as a byproduct of the present work we
have constructed the non--selfdual version of the Abelian solutions by departing from the Bogomol'nyi limit away from
the critical value of the Higgs selfinteraction strength. The resulting solutions exhibit the same properties of mutual
attraction and repulsion that are seen in the usual Abelian Higgs model itself~\cite{Jacobs:1978ch}. It appears that to
date this particular result has not appeared in the literature. After constructing these Abelian vortices, we
proceed to construct the fully non Abelian solutions, both for the value of the Higgs selfinteraction parameter for
which the Abelian embedding is selfdual, and, for other values of this parameter.

The paper is structured as follows: in the next section we present the model, impose rotational symmetry and discuss the
residual one-dimensional system.
Then in section {\bf 3} we carry out the numerical constructions and summarise our results in 
section {\bf 4}.

\section{The model}

\subsection{The action and the general Ansatz}

The CS-Higgs model on a $2+1$ dimensional Minkowski spacetime is described by the following Lagrangian
\be
\label{lag}
{\cal L}=\Omega_{\rm{CS}}+\mbox{Tr}\,D_{\mu}\F\,D^{\mu}\F-V(|\F|,\eta)
\ee
in which the CS density is
\be
\label{CS}
\Omega_{\rm{CS}}=\frac{\ka}{2}\,\vep^{\rho\mu\nu}\,\mbox{Tr}\,A_{\rho}\left(F_{\mu\nu}-\frac23\,A_{\mu}A_{\nu}\right)
\ee
and the symmetry breaking Higgs selfinteraction potential is that employed in \cite{Hong:1990yh,Jackiw:1990aw}
\be
\label{pot}
V=(4\la)^2\,\mbox{Tr}\,\F^2\,(\eta^2+\F^2)^2\,.
\ee
The dimensions of the various constants appearing above are, $[\eta]=L^{-1}$, $[\la]=L$ and $[\ka]=L^{-1}$, and the
index $\mu=0,i$, with $i=1,2$. The Lagrangian (\ref{lag}) usually enters the more complicated models as the basic
building block (see $e.g.$ \cite{Collie:2008za} and the references therein). Therefore  one can expect the basic features of its solutions to be
generic.

The static Hamiltonian of the Lagrangian \re{lag} is
\bea
{\cal H}_{\rm stat}&=&\frac12\,\left[\mbox{Tr}\left(D_0\F^2+D_i\F^2\right)+V(|\F|,\eta)\right] 
 =\frac12\,\left[\mbox{Tr}\left([A_0,\F]^2+D_i\F^2\right)+V(|\F|,\eta)\right]\,.\label{statHam}
\eea

We take the static "spherically" symmetric $SO(4)$ YM field in
$3$ spacetime ($i.e.$ $2$ Euclidean) dimensions, in one or other
chiral representation of $SO_{\pm}(4)$, such that our 
 Ansatz is expressed in terms of the representation matrices
\be
\label{sigmap}
\Sigma_{\al\beta}^{(\pm)}=-\frac{1}{4}\left(\frac{1\pm\gamma_{5}}{2}\right)
[\gamma_{\al} ,\gamma_{\beta}]\quad,\quad \al,\beta=1,2,3,4\ ,
\ee
$\gamma_{\al}=(\gamma_{i},\gamma_{M})$, with the index $M=3,4$, being
the gamma matrices in $4$ dimensions and $\gamma_{5}$, the chiral matrix.

Our rotationally symmetric Ansatz for the Higgs field $\F$ and the YM connection $A_{\mu}=(A_0,A_i)$ is
\bea
\nonumber
\F&=&-(\vep\f)^M\,n_j\,\Sigma_{jM}^{(\pm)}-
\f^{5}\,\Sigma_{34}^{(\pm)}\label{hp} \ ,
\\
A_0&=&-(\vep\chi)^M\,n_j\,\Sigma_{jM}^{(\pm)}-
\chi^{5}\,\Sigma_{34}^{(\pm)}
\label{a0p} \ ,
\\
\nonumber
A_i&=&\left[\left(\frac{\xi^M}{r}\right)(\vep\hat x)_i\,(\vep n)_j+
(\vep A_r)^M\,\hat x_i\,n_j\right]\Sigma_{jM}^{(\pm)}+
\left[A_r^{5}\,\hat x_i+\left(\frac{\xi^{5}+n}{r}\right)(\vep\hat
  x)_i\right]\Sigma_{34}^{(\pm)}\label{aip} \ ,
\eea
in which the functions $(\xi^M,\xi^{5})\equiv\vec\xi$,
$(\chi^M,\chi^{5})\equiv\vec\chi$ and $(A_r^M,A_r^{5})\equiv\vec A_r$ parametrise the YM connection in terms
of three sets of isotriplets $\vec\xi$, $\vec\chi$ and $\vec A_r$, the isotriplet $(\f^M,\f^{5})\equiv\vec\f$
parametrising the Higgs field. All four isotriplets depend only on
the $2$ dimensional spacelike radial variable $r$, $\vep$ is the two
dimensional Levi-Civita symbol, and $n_i=(\cos n\vf,\sin n\vf)$ is the unit vector encoded with the winding (vortex)
number $n\geq 1$ in the $(x_1,x_2)$ plane (with $r^2=x_1^2+x_2^2$).

Having stated the Ansatz \re{hp}  in terms of the gamma matrices in $4$
dimensions, we adopt henceforth the simpler labeling $\vec\xi=(\xi^M,\xi^{3})$,
$\vec\chi=(\chi^M,\chi^{3})$, $\vec\f=(\f^M,\f^{3})$, and $\vec A_r=(A_r^M,A_r^{3})$, with $M=1,2$ now.

\subsection{The residual system and a consistent truncation}
The parametrisation used in \re{hp}  results in $SO(3)$ gauge
covariant expressions for the YM curvature $F_{\mu\nu}=(F_{ij},F_{i0})$ and the components of the covariant derivative
of the Higgs field $D_{\mu}\F=(D_{i}\F,D_{0}\F)$, expressed exclusively in terms of the covariant derivatives of the
three triplets $\vec\xi=(\xi^M,\xi^{3})$, $\vec\chi=(\chi^M,\chi^{3})$ and $\vec\f=(\f^M,\f^{3})$, these covariant
derivatives in the residual one dimensional space being defined as
\be
\label{covp}
D_r\xi^a=\pa_r\xi^a+\vep^{abc}\,A_r^b\,\xi^c\quad,\quad
D_r\chi^a=\pa_r\chi^a+\vep^{abc}\,A_r^b\,\chi^c\quad,\quad
D_r\f^a=\pa_r\f^a+\vep^{abc}\,A_r^b\,\f^c~.
\ee
That the residual one dimensional system of fields resulting from
the imposition of this symmetry is entirely expressed in terms of the $SO(3)$ covariant quantities \re{covp} is a
consequence of the consistency of Ansatz \re{hp}, which has been verified explicitly.

The Euler--Lagrange equations arising from the variation of the gauge field are
\bea
\frac{\ka}{2}\,\vep_{ij}F_{ij}+\left[\F,\left[A_0,\F\right]\right]&=&0,
\label{gauss}
\\
\nonumber
\ka\,\vep_{ij}F_{j0}-\left[\F,D_i\F\right]&=&0,
\label{biot}
\eea
the first of which being the Gauss Law equation. The Higgs equation is
\be
\label{higgs}
D_iD_i\F-\left[A_0,\left[A_0,\F\right]\right]-
\la\eta^2\,\F\,\left(\eta^2+\F^2\right)\left(\eta^2+3\,\F^2\right)=0\,.
\ee
With the notation \re{covp}, the gauge field
equations \re{gauss} reduce to the following set of ordinary differential equations:
\bea
\frac{\ka}{2r}\,D_r\vec\xi&=&-
\left[|\vec\f|^2\,\vec\chi-(\vec\f.\vec\chi)\,\vec\f\right]
\label{ode1} \ ,
\\
\nonumber
\frac{\ka\,r}{2}\,D_r\vec\chi&=&-
\left[|\vec\f|^2\,\vec\xi-(\vec\f.\vec\xi)\,\,\vec\f\right]
\label{ode2} \ ,
\eea
together with the constraint equation
\be
\label{constr}
\vec\f\times D_r\vec\f=\frac{\ka}{2r}\,\vec\xi\times\vec\chi\,.
\ee
The Higgs equation reduces to
\be
\label{odeh}
D_r(r\,D_r\vec\f)-\frac1r\,\left[|\vec\xi|^2\,\vec\f-(\vec\f.\vec\xi)\,\,\vec\xi\right]
+r\,\left[|\vec\chi|^2\,\vec\f-(\vec\f.\vec\chi)\,\,\vec\chi\right]
-\la^2\,r\,(\upsilon^2-|\vec\f|^2)(\upsilon^2-3|\vec\f|^2)\,\vec\f=0\,.
\ee
In \re{odeh}, we have rescaled the VEV $\eta$ as $\eta=\frac12\, \upsilon$ to simplify
the expression. We note that the winding number $n$ in the Ansatz does not appear explicitly in the equations of
motion \re{ode1}  and \re{odeh}, nor in the contraint equation \re{constr}. It will appear only in the
boundary value of the function $\xi^3$ at the origin.

That \re{constr} is a constraint equation is easily verified by acting on it
with $D_r$ and employing the other three field equations.

In what follows we shall set the triplet of functions $\vec A_r=0$. That this is a consistent truncation is obvious
since there is no curvature in one dimension.

Substituting \re{hp}  in \re{lag} yields the following reduced one dimensional Lagrangian
\bea
L&=&\frac{\ka}{2}\left[\vec\chi\cdot\vec\xi_r-(\vec\xi\cdot\vec\chi_r+n\,\chi^3_r)\right]
+r\,|\vec\f\times\vec\chi|^2-\left(r\,|\vec\f_r|^2+\frac1r|\vec\f\times\vec\xi|^2\right)-\la^2r\,|\vec\f|^2
(\upsilon^2-|\vec\f|^2)^2\,.\label{redlag}
\eea
The equations of motion \re{ode1} and \re{odeh}, with all covariant derivatives replaced by the
ordinary derivatives $D_r\vec\f\to\frac{d\vec\f}{dr}\equiv{\vec\f}_r$, {\it etc.}, follow from the variation of \re{redlag}
with respect to the three triplets $\vec\chi$ , $\vec\xi$ and $\vec\f$ respectively.

It turns out that the Ansatz \re{a0p} can be consistently truncated further by setting
\be
\label{trunc0}
\vec\xi=(c\,k^M\,,\,a)\quad,\quad\vec\chi=(d\,k^M\,,\,b)\quad,\quad\vec\f=(\upsilon
h\,k^M\,,\upsilon \,g) \ ,
\ee
where $k^M$ is a constant {\it unit length} $2-$vector~\footnote{Using different constant {\it unit length} $2-$vectors
in $\vec\xi$, $\vec\chi$ and $\vec\f$ does not lead to a consistent truncation.}. We are thus left with only six radial
functions $a(r)$, $b(r)$, $c(r)$, $d(r)$, $h(r)$ and $g(r)$
resulting in the truncated version of the reduced Lagrangian \re{redlag}
\bea
L_{\rm trunc}&=&-\frac\ka 2[(a\,b_r-b\,a_r)+(c\,d_r-d\,c_r)+n\,b_r]+\upsilon^2\,r\,(bh-dg)^2\nonumber\\
&&-\upsilon^2\left[r(h_r^2+g_r^2)+\frac1r(ah-cg)^2+\upsilon^4\,\la^2\,r\,(h^2+g^2)[1-(h^2+g^2)]^2\right]
\label{trucredlag}
\eea
leading to the static energy density functional resulting from \re{trucredlag}
\bea
H_{\rm trunc}&=&\frac12\,\upsilon^2\left[r\,(bh-dg)^2
+r(h_r^2+g_r^2)+\frac1r(ah-cg)^2+\upsilon^4\,\la^2\,r\,(h^2+g^2)[1-(h^2+g^2)]^2\right]\,.\label{trucredham}
\eea

\subsection{The Abelian case}
It is natural at this stage to isolate the Abelian embedding of this system, which will play an essential role in
the construction of the non Abelian solutions. The Abelian embedding results from the truncation
\[
c=d=g=0  \ ,
\]
for which the constraint \re{constr} is identically satisfied.

The remarkable feature of this Abelian case is that the Gauss Law equation is an algebraic equation enabling
the elimination of the electric component $A_0$ of the Maxwell field $A_{\mu}=(A_i,A_0)$. This results in the
reduction of the static Hamiltonian becoming identical with that of the usual Abelian Higgs model, subject to
an additional {\it proviso} restricting the symmetry breaking Higgs selfinteraction potential to be of the form
\re{pot}, namely the natural case chosen in \cite{Hong:1990yh,Jackiw:1990aw}.

The Gauss Law equation now reduces to
\be
\label{gaussU1}
b=-\frac{\ka}{2\upsilon^2h^2}\,\frac{a_r}{r}\,,
\ee
leading to the reduced one dimensional Lagrangian
\be
\label{redlagu1}
L_{U(1)}=\frac{\ka}{2}\left[b\,a_r-(a+1)\,b_r\right]+\upsilon^2\,r\,b^2\,h^2-
\upsilon^2\left(r\,h_r^2+\frac{a^2\,h^2}{r}\right)-\la^2\,\upsilon^6\,r\,h^2(1-h^2)^2   \ .
\ee
The Gauss Law constraint \re{gaussU1}, can now also be derived from the variation of \re{redlagu1} with respect to $b(r)$.

Next, using integration by parts, we replace the term $(a+1)\,b_r$ in \re{redlagu1} by $b\,a_r$, whence \re{redlagu1}
can be expressed as
\be
\label{redlagu2}
L_{U(1)}=-\upsilon^2\left\{\left[\frac{\ka^2}{4\upsilon^4}\frac{a_r^2}{rh^2}
+\la^2\,\upsilon^4\,r\,h^2(h^2-1)^2\right]+\left(r\,h_r^2+\frac{a^2\,h^2}{r}\right)\right\}\equiv
-H_{\rm {stat}}   \ .
\ee
To establish the topological lower bound of $H_{\rm {stat}}$ defined by \re{redlagu2}, we consider the two inequalities
\bea
\left(\sqrt{r}\,h_r\mp\frac{a\,h}{\sqrt{r}}\right)^2&\ge&0 \,,\nonumber\\
\left[\frac{\ka}{\upsilon^2}\,\frac{a_r}{2\sqrt{r}\,h}\mp\la\,\upsilon^2\,
\sqrt{r}\,h\,(h^2-1)\right]^2&\ge&0\,,\nonumber
\eea
which lead to the final inequality
\be
\label{lb11}
H_{\rm {stat}}\ge \pm \upsilon^2 \left[a(h^2-1)_r+\ka\la\,a_r(h^2-1)\right]\,.
\ee
Now for this inequality to present a topological lower bound on the
energy, the right hand side of \re{lb11} must be a total derivative. This is
only possible if we choose the constants subject to
\be
\label{choice1}
\ka\,\la=1\,.
\ee
Saturating \re{lb11} with \re{choice1} yields precisely the Bogomol'nyi equations satisfied by the selfdual
Chern--Simons vortices of \cite{Hong:1990yh,Jackiw:1990aw}, which is relevant here because our numerical analysis
will depart from these vortices, first to the non--selfdual case analogous to the corresponding Abelian Higgs
vortices~\cite{Jacobs:1978ch}, and finally to the non Abelian vortices.

\subsection{The global charges}
On the question of the magnetic and electric fluxes, the situation is as follows. For the generic Abelian case, the
operative equations are the Maxwell equations, and of these the Gauss Law equation is \cite{Jackiw:1990aw}
\be
\label{gausslaw}
\nabla\cdot E-\ka\,B=\rho
\ee
where $E_i=F_{i0}$ and $B=F_{12}=\frac12\vep_{ij}F_{ij}$. The density $\rho$ is the $0-$th component of the $U(1)$
current expressed in terms of the complex Higgs field such that its volume integral over $d^2x$ is the electric flux.
The volume integral of
the divergence term $\nabla\cdot E$, after converting it to a 'surface' (line) integral, vanishes. This can be seen
readily from the results of the foregoing asymptotic analysis. As a result of the vanising of $\nabla\cdot E$ in
\re{gausslaw} for our model, it is clear that the magnetic flux $\int B\,d^2x$ is inversely related to the electric flux.

The only topological charge in this system is the magnetic flux, and hence also the electric flux. These are defined in
the context of the Abelian subsystem of the $SU(2)$ model at hand, and these respective charges
are the only global quantities pertaining to the solutions studied. The nonabelianness therefore has no influence in
this sector. There is however the angular momentum, or the spin, of these solutions, which presents another
global quantity characterising our solutions. Unlike the magnetic and electric charges however, which are not influenced
by the nonabelianess, the angular dependence is indeed dependent on the gauge group.

The angular momentum density is
\be
\label{densJ}
{\cal J}=T_{\vf 0}=(x\,\vep)_i\,T_{i0}=(\hat x\,\vep)_i\,r\,\mbox{Tr}\,D_i\F\,D_0\F\,,
\ee
which, for the fields subjected to the Ansatz \re{hp}, and further truncated
according to \re{trunc0}, readily yields,
\bea
{\cal J}&=&\frac12\,\upsilon^2(\vec\f\times\vec\chi)\cdot(\vec\f\times\vec\xi)\nonumber\\
&=&\frac12\,\upsilon^2(a\,h-c\,g)(b\,h-d\,g)\label{calJ}\,.
\eea
Remarkably enough, the total angular momentum, which is given by the integral
\be
\label{J}
J=2\pi\,\int\,{\cal J}\,r\,dr\,,
\ee
can be expressed as a difference of two boundary integrals\footnote{A similar property  of the total angular momentum has been noticed in $3+1$
dimensions for various models with gauged fields, see $e.g.$ \cite{Radu:2008pp} and the references there.
However, in contrast to the model here, for all known $d=4$ cases, 
the contribution to $J$ of the inner boundary term vanishes.}, (thus in this one dimensional case it reduces to the
integral of a total derivative). It follows from the field equations (\ref{gauss}), that the integral \re{J} can be
written as
\be
\label{J1}
J=-\ka\,\frac{\pi}{2}\,\int_{0}^{\infty}\,\vec\xi \cdot D_r\vec\xi dr=
-\ka\,\frac{\pi}{4}\,(|\vec\xi(\infty)|^2-|\vec\xi(0)|^2)\,
=\ka\,\frac{\pi}{4}\,(n^2-p_1^2)\,.
\ee
Note that $p_1$ appearing in \re{J1} is a asymptotic parameter appearing in \re{alpha} below, and that the Abelian
embedding solution is consistent with the value $p_1=0$ of this parameter. Thus in this limit, the expression \re{J1}
coincides with the angular momentum given in \cite{Hong:1990yh,Jackiw:1990aw}, $i.e.$  $J=\ka\,{\pi n^2}/{4}$,
which holds for both selfdual and non-selfdual solutions. It is clear that the angular momentum
of the non Abelian vortices differs essentially from that of its Abelian counterparts and that it can only be evaluated
numerically, which will be described in the following section.

We further define the energy of the solutions as the integral of
\re{trucredham}, namely,
\be
\label{energy}
E = 2\pi\,\int\, r T_{00} \,dr= 2\pi\,\int\, H_{\rm trunc} \,dr \,.
\ee

\section{Numerical results}
\label{numerics}

\subsection{General features}
Although an analytic or approximate solution appears to be
intractable
here we present arguments 
for the existence of  nontrivial solutions of the field equations \re{ode1}-\re{odeh}.

For the full non Abelian system,  the asymptotic expansion of the solutions near the origin is
\begin{eqnarray}
\nonumber
 a(r)&=&-n-\frac{ d_n^2}{ 4(n+1)}(2+\frac{4b_0 h_h^2 \upsilon^2}{d_n^2\kappa})r^{2n+2}+O(r^{2n+4})\,,\label{change}
\\
\nonumber
b(r)&=&b_0+\frac{h_n^2 \upsilon^2}{\kappa } r^{2n}+O(r^{2n+2}),
\\
\nonumber
c(r)&=&-\frac{d_n}{ (n+2)} ( b_0+\frac{d_n^2\kappa}{2 h_n^2 \upsilon^2})r^{n+2} +O(r^{n+4}),
\\
d(r)&=& d_n r^n +O(r^{n+2}),
\\
\nonumber
h(r)&=& h_n r^n +O(r^{n+2}),
\\
\nonumber
g(r)&=&-\frac{d_n\kappa}{2h_n \upsilon^2}-\upsilon^2\lambda^2\frac{d_n\kappa}{8h_n}
\left( 1-\frac{3d_n^2\kappa^2}{4h_n^2 \upsilon^4} \right)
\left( 1-\frac{ d_n^2 \kappa^2 }{4h_n^2 \upsilon^4} \right)
r^2 +O(r^4) \, .
\end{eqnarray} 
All higher order terms in this expansion are fixed by the
 coefficient $b_0,d_n,h_n$.
Thus, at the origin one uses the following set of boundary conditions 
\begin{eqnarray}
\label{c1}
a|_{r=0}=-n,~~\partial_r b|_{r=0}=0,~~ c|_{r=0}= d|_{r=0}= h|_{r=0}=0~,\partial_r g|_{r=0}=0~.
 \end{eqnarray}
At infinity, the finite energy requirements impose
\begin{eqnarray}
g= \cos \alpha,~~h= \sin \alpha,~~
a=p_1\cos \alpha,~~c=p_1\sin \alpha,~~
d=p_2\sin \alpha,~~b=p_2\cos \alpha,~~\label{alpha}
\end{eqnarray}
where $\alpha,p_1,p_2$ are arbitrary constants fixed by numerics.
Physically, $p_1$ and $p_2$ corresponds to the asymptotic amplitudes of the effective scalars $\vec \chi$ and $\vec \xi$,
respectively, while $\alpha$ somehow characterises the angle of the Higgs field with respect to
the Abelian solution, since $\alpha=\pi/2$ in that limit.

Let us now concentrate on the numerical resolution of equations (\ref{ode1}) 
and (\ref{odeh}), together with the constraint given by equation (\ref{constr}). 
In order to do so, we employ a collocation method for boundary-value ordinary
differential equations\footnote{Some of the solutions were also constructed by  using a standard Runge-Kutta
ordinary  differential  equation solver. In this approach we 
evaluate the  initial  conditions  at $r=10^{-5}$ for  global  tolerance  $10^{-12}$,
adjusting  for shooting parameters $h_n,b_0$ and  integrating  towards  $r\to\infty$.
We have noticed a very good agreement between the results obtained with these two different methods. 
The accuracy of the solutions was also monitored by computing a 
virial relation satisfied by the system (\ref{trucredlag}).}, equipped with an adaptive mesh selection procedure
\cite{colsys}.
Typical mesh sizes include $10^3-10^4$ points. The solutions have a relative accuracy of $10^{-7}$.

It is worth noticing that equations (\ref{ode1})-(\ref{odeh}) may be rescaled by
\be
b \to \frac{2 \upsilon^2}{\kappa} b \quad,\quad d \to \frac{2 \upsilon^2}{\kappa} d
\quad,\quad  r \to \frac{\kappa}{2 \upsilon^2} r \quad,\quad  \lambda  \to
\frac{\nu}{\kappa} \, , \label{rescaling}
\ee
such that dependence on $\kappa$ and $\upsilon$ disappears, remaining just a
dependence on $\nu$, which encodes the Higgs self-coupling parameter.
For that reason, without loss of
generality,  we will assume in what follows that $\kappa=2 \upsilon^2$ and $\upsilon=1$ and
$\lambda$ will be rewritten as $\nu/2$. 
With this convention, the Abelian solutions approach the selfdual limit for $\nu=1$.
For numerical reasons we further introduce a compactified radial coordinate defined by $\bar r = r/(1 + r)$. 

After a detailed analysis of the equations, one finds that for a fixed integer
$n$ and a non-vanishing real $\nu$, the regular solutions to
(\ref{ode1})-(\ref{odeh}) depend on just one numerical free parameter. 
We have
chosen it to be $p_2$ so the two remaining constants in (\ref{alpha}), namely, $p_1$ and $\alpha$, 
are not free but are given by numerics. For that reason, in our numerical scheme the boundary conditions at infinity
were chosen to be
\be
d - p_2 h = 0 \quad,\quad b h -d g = 0 \quad,\quad g^2 + h^2 = 1  \, . \label{BC_infty}
\ee
To summarize, the numerical solutions are constructed by using the boundary conditions (\ref{c1}), (\ref{BC_infty}) 
with the following input parameters: the winding number $n$, 
the Higgs self-coupling constant $\nu$ and the asymptotic value $p_2$ of the electric non Abelian
potential\footnote{It is interesting to remark that
these are also the usual input parameters for the dyonic Yang-Mills-Higgs (YMH) solutions in $3+1$ dimensions,
see $e.g.$ \cite{Hartmann:2000ja}.}.

Our procedure to generate non-Abelian solutions in the $\{n,\nu,p_2\}$
parameter space was as follows: for fixed integer $n$, one starts from the
corresponding selfdual Abelian solution ($\nu=1$ and $p_2=0$), which
corresponds to the  solution in \cite{Hong:1990yh}, \cite{Jackiw:1990aw};  moving
$\nu$ from $1$ while keeping $p_2=0$ one generates non-selfdual Abelian
solutions; moving $p_2$ from zero\footnote{As we will show later for large
  values of $\nu$ there are
  non-Abelian solutions with $p_2=0$ in addition to the Abelian ones.} while keeping $\nu=1$ one generates
non-Abelian solutions with $\nu=1$;  finally, the  general solutions are found when moving both $\nu$ and $p_2$.
Also, nontrivial solutions are likely to exist for any value of the winding number $n$; however, in practice,
numerics becomes more involved with increasing $n$. Similarly, we have
found non-Abelian solutions for arbitrarily large values of $\nu$.

In Figure~1 we show the functions $a$, $b$, $c$, $d$, $g$, and $h$
for a typical non-Abelian solution (with $\nu=2.0$ and $p_2=0.7$) for $n=1$.
The remarkable non-Abelian nature of these solutions is inferred from the
significant deviation from zero of the functions $c$, $d$, and $g$.

\setcounter{fixy}{1}

   \begin{fixy} {-1}
\begin{figure}[h!]
\parbox{\textwidth}
{\centerline{
\mbox{
\epsfysize=10.0cm
\includegraphics[width=70mm,angle=-90,keepaspectratio]{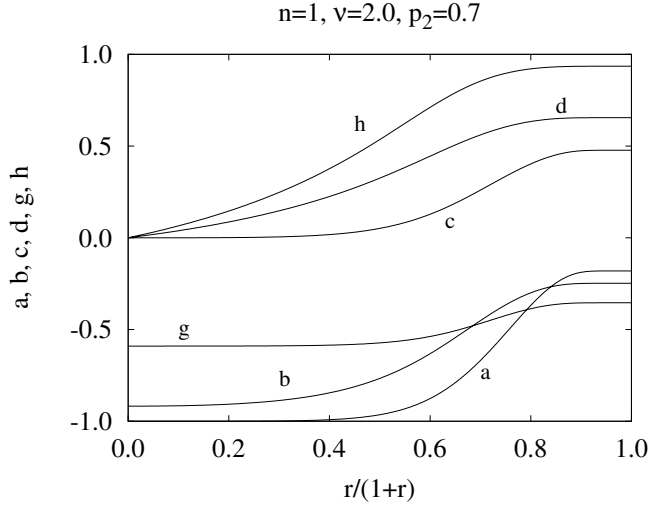}
}}}
\caption{A typical non-Abelian solution with $\nu=2.0$ and
  $p_2=0.7$ is shown for $n=1$.
}
\end{figure}
     \end{fixy}

\subsection{The energy of the solutions}

Let us analyze the energy of the solutions. In Figure~2 we show the energy per vortex number of the Abelian solutions
($p_2=0$) versus $\nu$ for $n=1$, $2$, $3$. In the selfdual limit ($\nu=1$)
the curves coincide. Below $\nu=1$ the energy per vortex number decreases with
$n$ for fixed $\nu$ whilst above $\nu=1$ it increases monotonically with
$n$. This can be interpreted by saying that force between vortices is
attractive below the selfdual limit ($\nu=1$), while it is repulsive above
$\nu=1$. We also include for comparison the energy per vortex number of non-Abelian
solutions with $p_2=0.5$. In that case there is no fixed selfdual limit where
all the curves merge but curves cross in pairs at several values of $\nu$. 

   \begin{fixy} {-1}
\begin{figure}[h!]
\parbox{\textwidth}
{\centerline{
\mbox{
\epsfysize=10.0cm
\includegraphics[width=70mm,angle=-90,keepaspectratio]{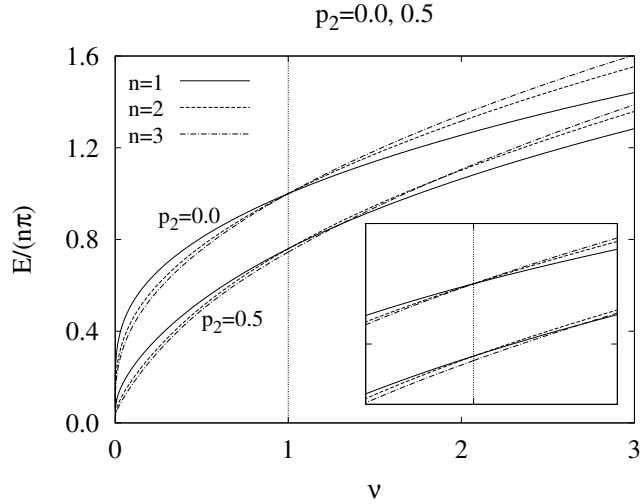}
}}}
\caption{Energy per vortex number $E/n$ versus the Higgs self-coupling
  constant $\nu$  for Abelian ($p_2=0.0$) and non-Abelian
  ($p_2=0.5$) solutions with $n=1$, $2$, $3$ .
In the Abelian case, the curves meet for $\nu=1$, the selfdual limit.}
\end{figure}
     \end{fixy}

The effect of non-Abelianity is exhibited in  Figures~3. In Figure~3a the energy per vortex
number is presented as a function of the parameter $p_2$ for several values of
$\nu$ for $n=1$, $2$, $3$. Starting from the corresponding Abelian solution
($p_2=0$) we move\footnote{Recall that the energy is an even function of
  $p_2$  and thus similar solutions exist when $p_2\to -p_2$.} $p_2$ for fixed $n$ and $\nu$. We observe numerically
that solutions exist only in the range\footnote{Physically, that means the amplitude of the electric potential at
infinity is always smaller than the asymptotic value of the Higgs field,
a feature present also in $3+1$ dimensional gauged Higgs models, see $e.g.$ \cite{Hartmann:2000ja}.}  $|p_2| \le 1$.

When the limit $p_2= \pm 1$ is approached the
solution tends to the trivial solution 
\be
a = -n \ ,\ b= -1 \ ,\ c=0 \ ,\ d=0 \ ,\ g = \mp 1 \ ,\ h =0 \ . \label{sol_lim} 
\ee
This limit requires some explanation. When $|p_2|
\to 1$ the sequence of solutions  tends to (\ref{sol_lim}) in a pointwise way. In fact, as we can see
from Figure~3a and subsequent ones, the energy for the limiting solution seems
to depend on $\nu$ and it is obviously nonvanishing. A naive
computation of the energy using (\ref{sol_lim}) however yields zero. The
explanation of this apparent contradiction may be understood by analysing
Figure~4. There the energy density ($\varepsilon = 2\pi H_{\rm trunc}$) of a sequence of solutions with $|p_2|
\to 1$ is presented. We observe that the energy density spreads and tends to
zero as $|p_2| \to 1$ but its integral remains finite. Moreover, if one
concentrates on functions $\{a,b,c,d,g,h\}$ for the sequence, one observes
that for any finite value of $r$ the sequence tends to the corresponding
limiting value although that is not necessarily true for the value at
infinity.


For small values of $\nu$ we find only one
solution for each value of $p_2$. However, for large values of
$\nu$ we find several solutions for the same value of $p_2$ (the larger $n$
is, the smaller $\nu$ needs to be) with different value of the energy, in general. In Figure~3a that is clearly seen
for $n=3, \nu=2$ where in the range $0.63 < |p_2| < 0.75$ three different non-Abelian
solutions coexist for each value of $p_2$.

Figures~3b and 3c show the energy per vortex number versus $p_1$ and $\alpha$,
respectively. It is clearly seen that the curves look much more complicated
when using $p_1$ or $\alpha$ as free parameter. Moreover, contrary to what
happens when using $p_2$ as free parameter, we do not observe any a priori
bound for $p_1$ and $\alpha$.

\begin{figure}[h!]
\parbox{\textwidth}
{\centerline{
\mbox{
\epsfysize=10.0cm
\includegraphics[width=70mm,angle=-90,keepaspectratio]{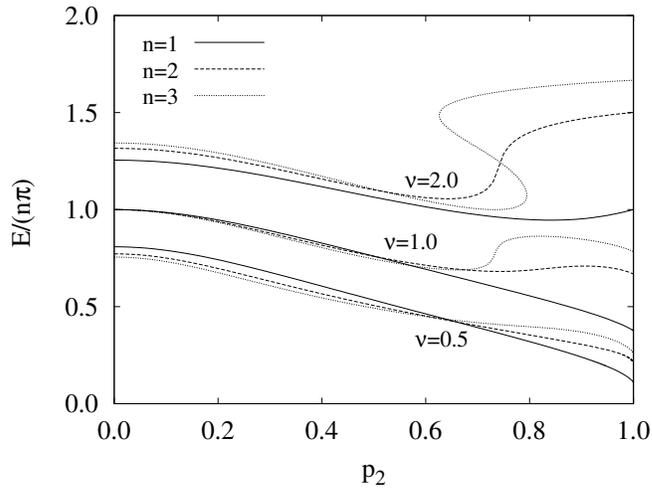}
}}}
\caption{Energy per vortex number $E/n$ versus the magnitude $p_2$ of the electric potential at infinity
for solutions with $n=1$, $2$, $3$ and several values of $\nu$. For $\nu=2,n=3$, one notices the existence
of several solutions with the same $p_2$.
}
\end{figure}

\begin{figure}[h!]
\parbox{\textwidth}
{\centerline{
\mbox{
\epsfysize=10.0cm
\includegraphics[width=70mm,angle=-90,keepaspectratio]{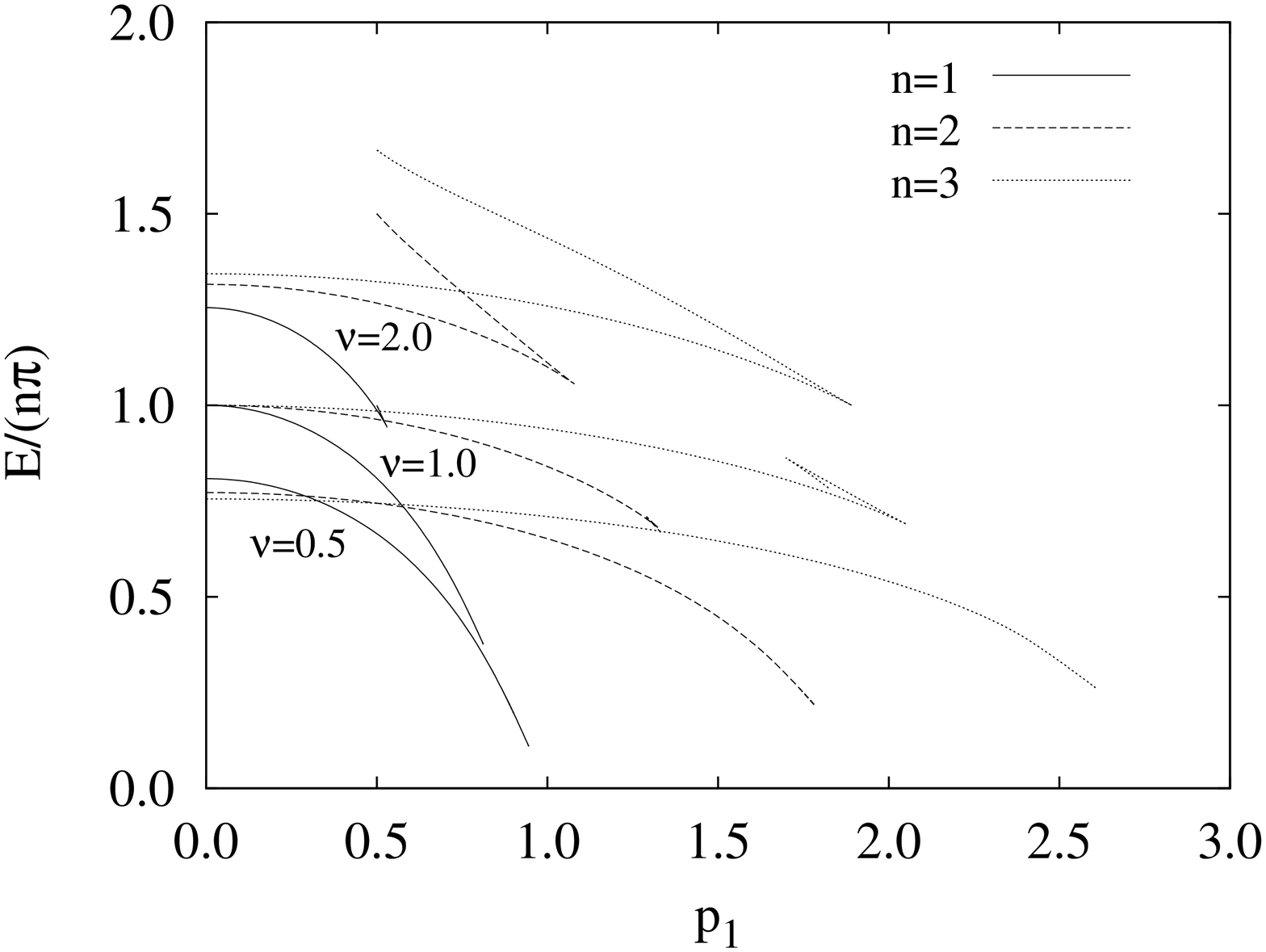}
}}}
\caption{Energy per vortex number $E/n$ versus $p_1$ for solutions with $n=1$,
  $2$, $3$ and several values of $\nu$.
}
\end{figure}

\begin{figure}[h!]
\parbox{\textwidth}
{\centerline{
\mbox{
\epsfysize=10.0cm
\includegraphics[width=70mm,angle=-90,keepaspectratio]{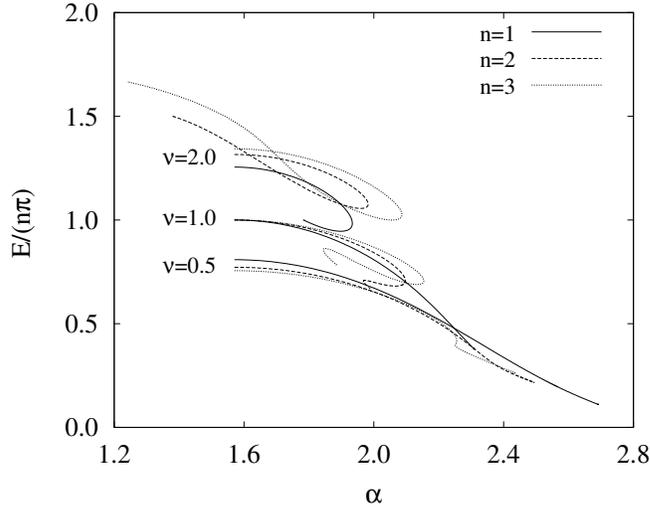}
}}}
\caption{Energy per vortex number $E/n$ versus $\alpha$ for solutions with $n=1$,
  $2$, $3$ and several values of $\nu$.
}
\end{figure}

\subsection{The angular momentum}

The angular momemtum of the solutions of Figure~3a is exhibited in Figure~5. Showing independence of $\nu$ for the Abelian case ($p_2=0$) we observe
that the angular momenta of solutions with the same value of $p_2$ but
different values of $\nu$ are different in general. As the masses of these
solutions also change with changing $p_2$ it is useful to consider the
behaviour of the angular momentum as a function of the energy.

\setcounter{fixy}{4}
   \begin{fixy} {-1}
\begin{figure}[h!]
\parbox{\textwidth}
{\centerline{
\mbox{
\epsfysize=10.0cm
\includegraphics[width=70mm,angle=-90,keepaspectratio]{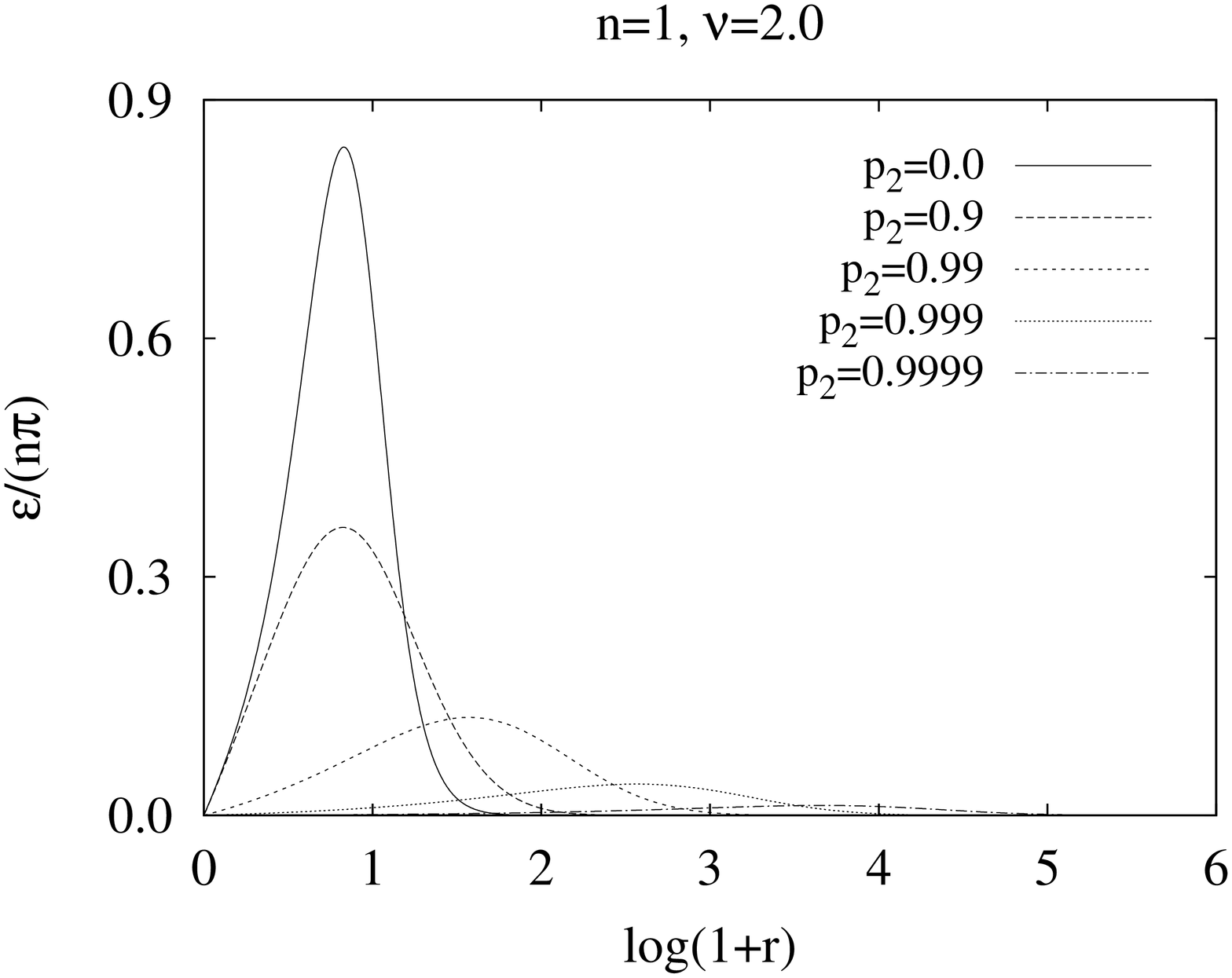}
}}}
\caption{Energy density $\varepsilon$ for a sequence of solutions with $|p_2|
  \to 1$ for $n=1$ and $\nu=2.0$.
}
\end{figure}
\end{fixy}

   \begin{fixy} {-1}
\begin{figure}[h!]
\parbox{\textwidth}
{\centerline{
\mbox{
\epsfysize=10.0cm
\includegraphics[width=70mm,angle=-90,keepaspectratio]{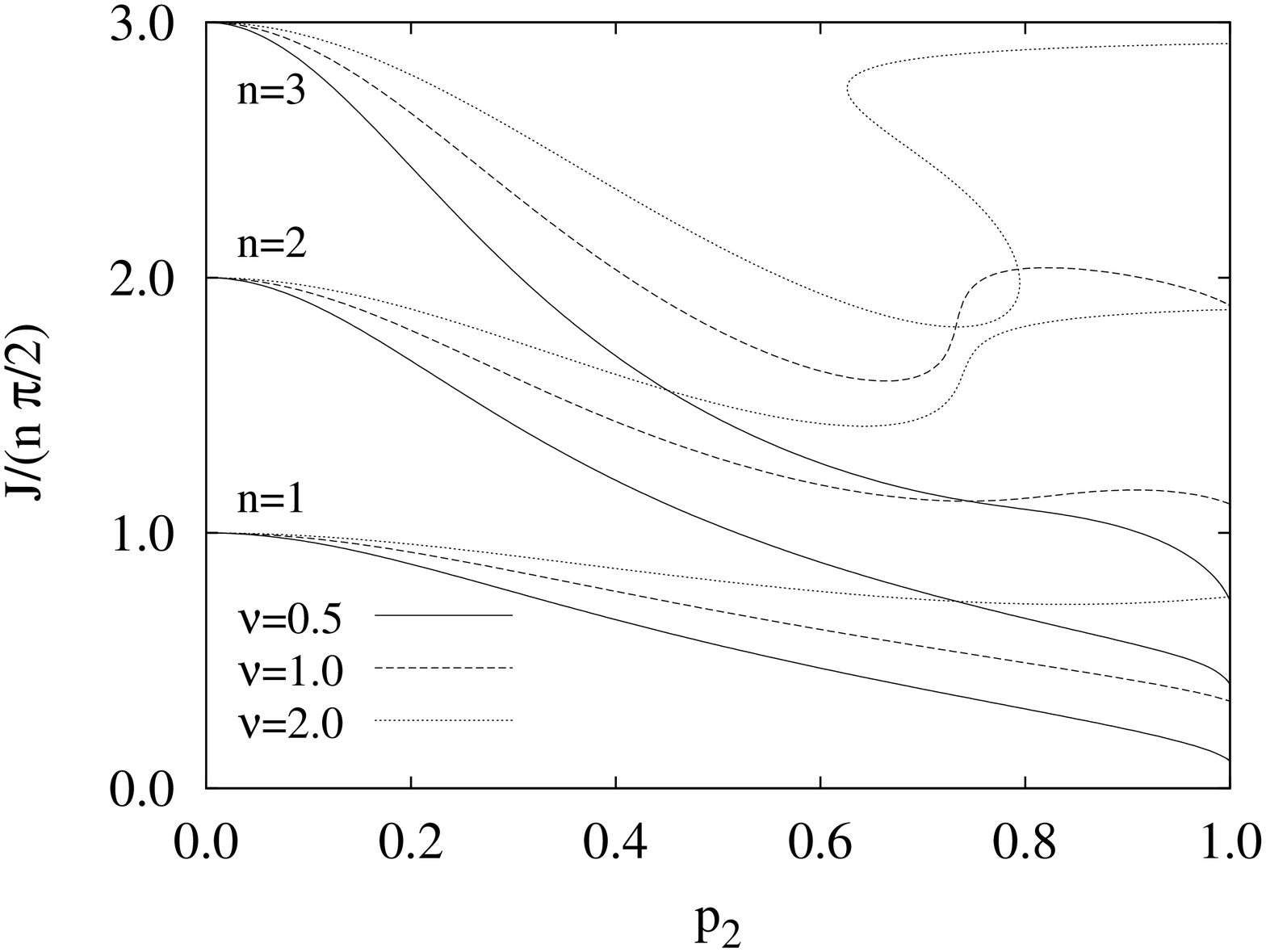}
}}}
\caption{Angular momentum per vortex number $J/n$ versus $p_2$ for the same set
  of solutions of Figure 3a.
}
\end{figure}
\end{fixy}

   \begin{fixy} {-1}
\begin{figure}[h!]
\parbox{\textwidth}
{\centerline{
\mbox{
\epsfysize=10.0cm
\includegraphics[width=70mm,angle=-90,keepaspectratio]{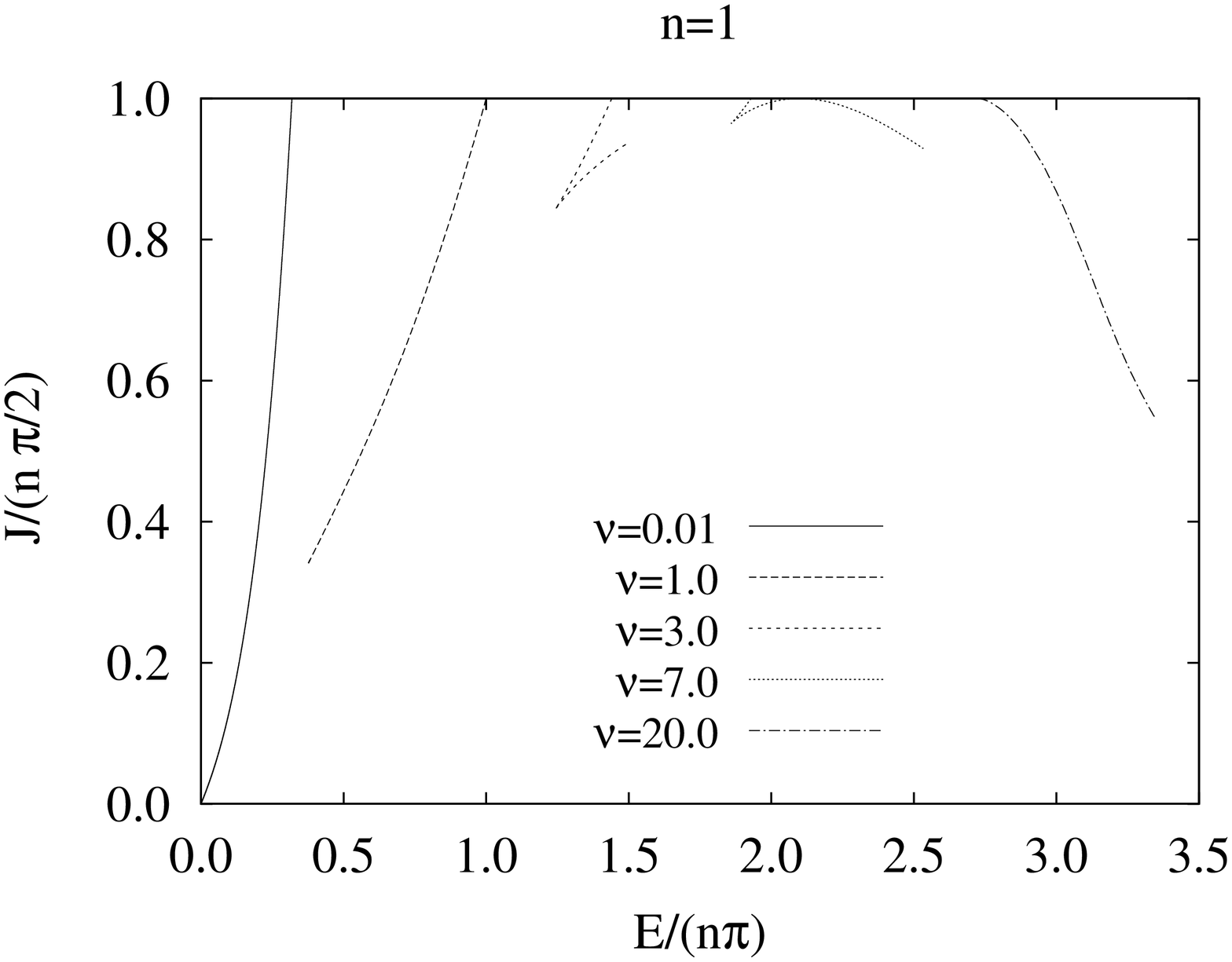}
}}}
\caption{Angular momentum per vortex number $J/n$ versus energy per vortex number $E/n$ for solutions with $n=1$ and several values of $\nu$.
}
\end{figure}
\end{fixy}

That is shown in Figure~6. For small values of $\nu$, the angular momentum $J$ is
an increasing function of the energy $E$. In fact, we observe that $J=0$ solutions
seem not to exist, except in the limit when the energy also vanishes. As $\nu$
is increased, the curves develop a kink but the angular momentum still remains
to be an increasing function of the energy. However, as $\nu$ is enlarged more,
one may find regions where the angular momentum becomes a decreasing function
of the energy. Finally, for very large values of $\nu$ the angular momentum
monotonically decreases with increasing energy. This strange effect has been
reported previously in other theories (for instance, in $d=3+1$
Einstein-Maxwell-dilaton theory \cite{Kleihaus:2003df}, associated to counterrotation).


   \begin{fixy} {-1}
\begin{figure}[h!]
\parbox{\textwidth}
{\centerline{
\mbox{
\epsfysize=10.0cm
\includegraphics[width=70mm,angle=-90,keepaspectratio]{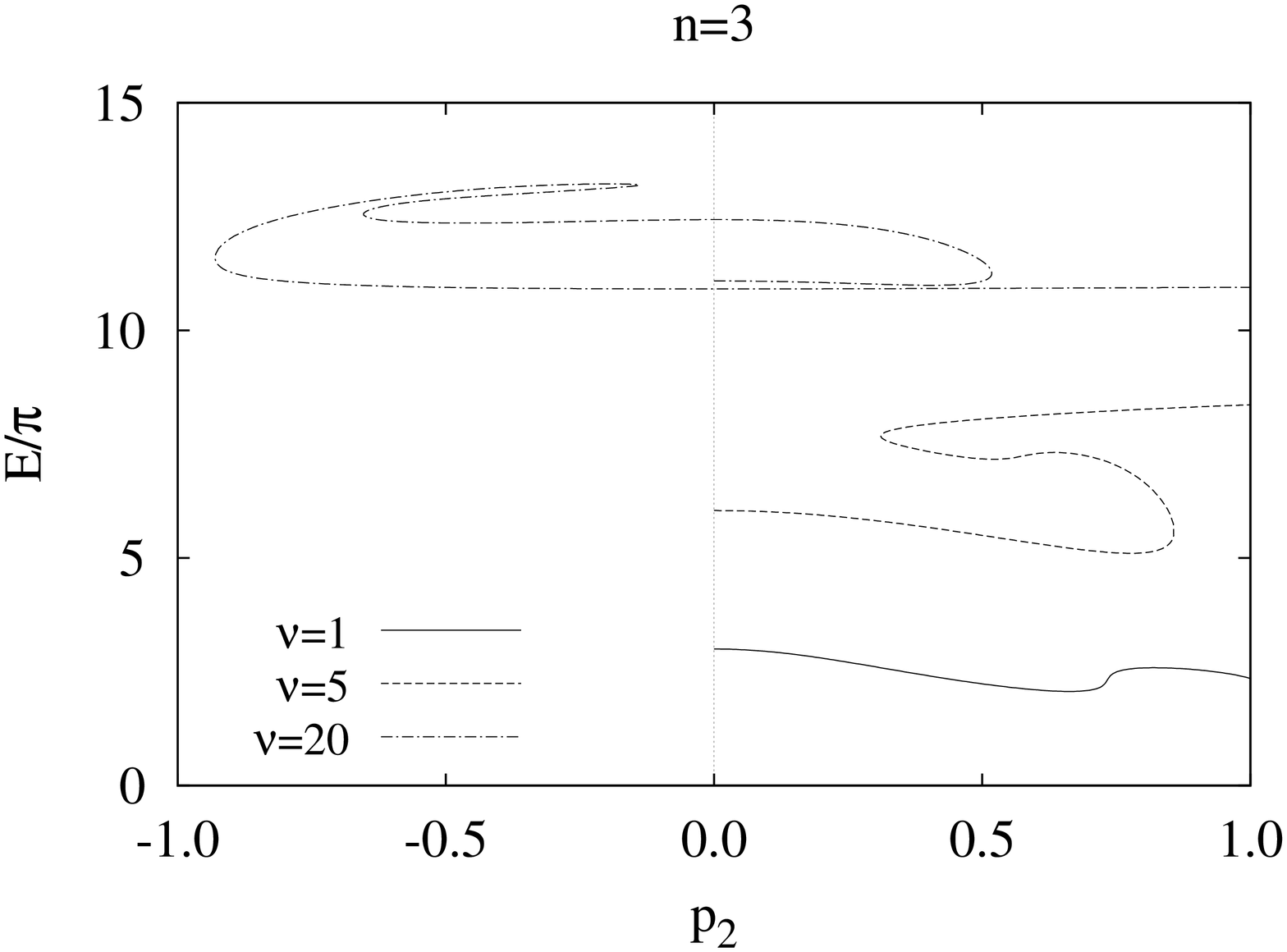}
}}}
\caption{Energy $E$ versus $p_2$ for solutions with $n=3$ and several values of $\nu$.
}
\end{figure}
\end{fixy}

\subsection{The issue of $p_2=0$ solutions}
Unexpectedly, the multiplicity of solutions in $p_2$ we observe in Figure~3a as a function of $\nu$,
 happens also at $p_2=0$ (although it cannot be seen there). Thus, the
condition $p_2=0$  ($i.e.$ $ |\chi| \to 0$ as $r\to \infty$) does not characterize Abelian
solutions~\footnote{This is unique feature of $d=2+1$ CSH theories. In the better known $d=4+1$ YMH case, $|A_0|=0$
at infinity implies a vanishing electric potential.}. These non-Abelian
$p_2=0$ solutions exist for any value of the vorticity $n$. In Figure~7 we show
this for solutions with $n=3$. It is clear how the curves get more complicated
as $\nu$ increases with the appearance of
non-Abelian $p_2=0$ solutions. Concerning this Figure~7 and the subsequent one
Figure~9, one should notice that only half of the complete curves are presented for
the sake of clarity. Since the energy is an even function of $p_2$ the complete
pictures would include also the mirror symmetric images, with respect to the $p_2=0$
line.

There is also one further feature one should mention. As general rule, there
are (almost always\footnote{We have observed however some small regions where
  that is not the case.}) non-Abelian solutions with lower energy than the connected
Abelian solution for fixed values of $n$ and $\nu$. This fact may take
consequences on the stability of Abelian solutions.

Let us analyse further these $p_2=0$ solutions. In Figure~8 we exhibit the
energy $E$ of $p_2=0$ solutions for $n=1$. The picture is similar for any
other value of $n$ (although there are more kinks as $n$ increases). Plotting
the Abelian branch of solutions we observe there is a value of $\nu$
($\nu \simeq 75.0, 54.2, 55.6$ for $n=1, 2, 3$, respectively) at which the
non-Abelian $p_2=0$ branch branches off (represented by a large dot in
Figure~8). Following the non-Abelian branch we observe it crosses the Abelian
branch at another value of $\nu$. That means it is possible to
find two different solutions with the same values of $\{n, \nu, p_2, E \}$.
Above that value we observe the existence of non-Abelian $p_2=0$
solutions with energy lower than the corresponding energy of their Abelian counterparts.


   \begin{fixy} {-1}
\begin{figure}[h!]
\parbox{\textwidth}
{\centerline{
\mbox{
\epsfysize=10.0cm
\includegraphics[width=70mm,angle=-90,keepaspectratio]{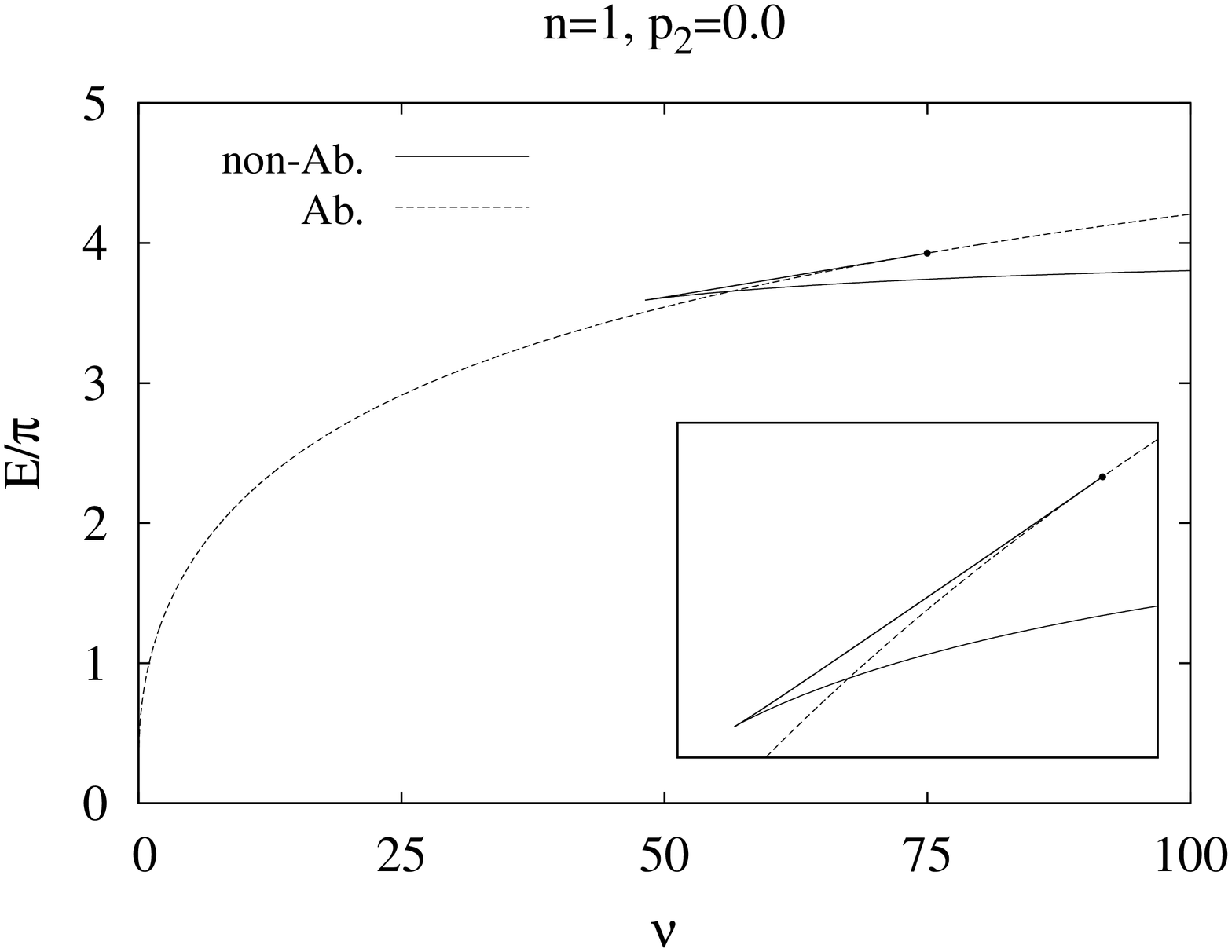}
}}}
\caption{Energy $E$ versus the Higgs self-coupling parameter $\nu$ for $p_2=0$ solutions with $n=1$.
For a range of $\nu$, the nonabelian solutions have a lower energy than their abelian counterparts. 
}
\end{figure}
\end{fixy}

   \begin{fixy} {-1}
\begin{figure}[h!]
\parbox{\textwidth}
{\centerline{
\mbox{
\epsfysize=10.0cm
\includegraphics[width=70mm,angle=-90,keepaspectratio]{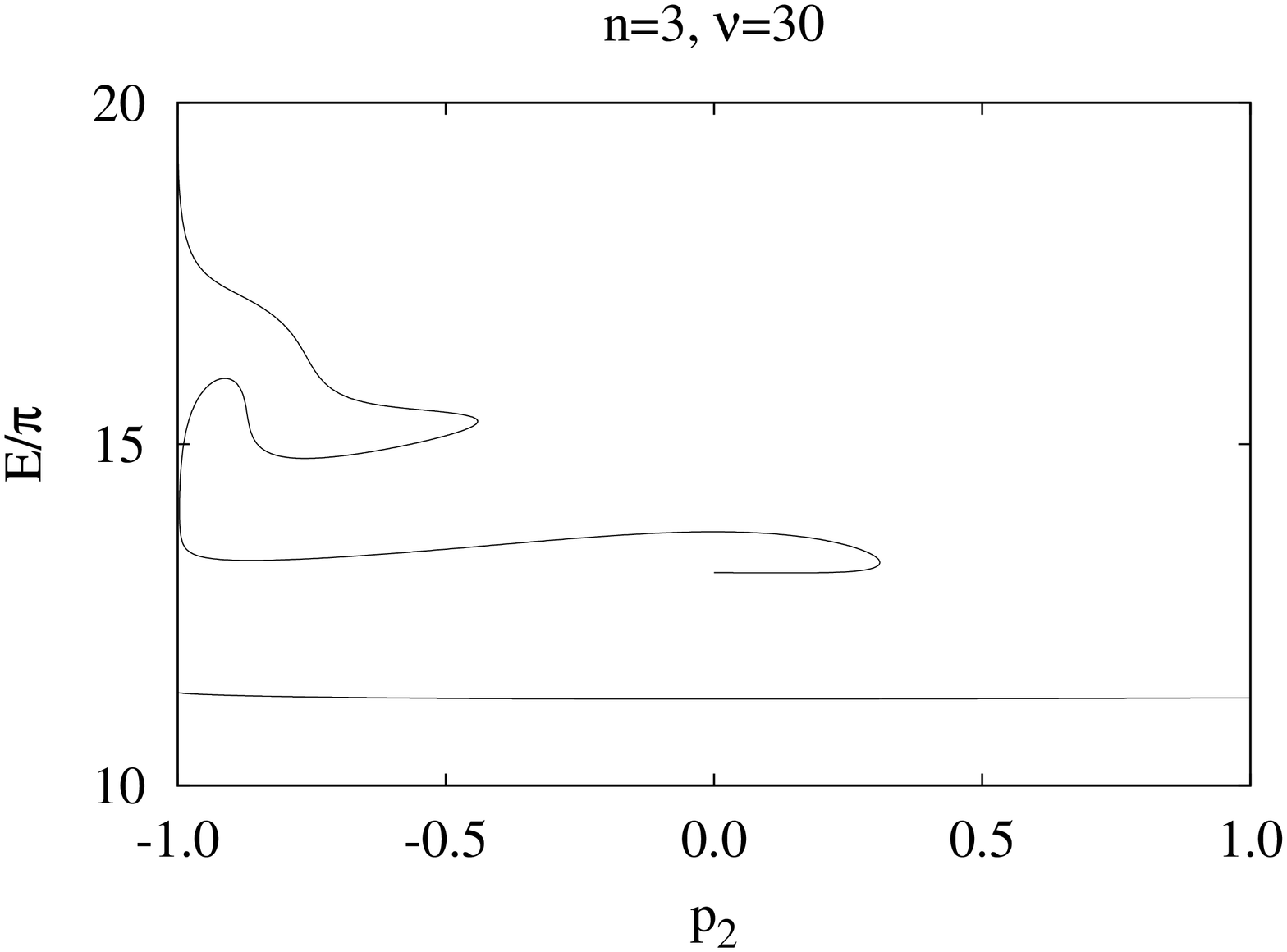}
}}}
\caption{Energy $E$ versus $p_2$ for $n=3$, $\nu=30$ solutions. A 
  branch disconnected from the Abelian solution (starting point at $p_2=0$) is shown .
}
\end{figure}
\end{fixy}

\subsection{Disconnected branches}

Even more strange situations may happen for large values of the Higgs
self-coupling constant $\nu$. There we have found numerical evidence for the existence of disconnected
non-Abelian branches. In general, all these non-Abelian solutions reported above
are obtained by continuous variation of the free parameter $p_2$ in the range
$|p_2| \le 1$. However, by fine-tuning $\nu$ and $p_2$ it is possible to
reach a region of the parameter space where the corresponding Abelian solution
cannot be reached by just moving $p_2$, $i.e.$, one can move $p_2$ from -1 to 1
(passing through 0) without reaching the Abelian solution. An example of these
disconnected branches is presented in Figure~9 for $n=3$ and $\nu=30$.

\section{Summary and discussion}
\label{conclusions}
We have constructed non Abelian vortices in a $SU(2)$ Chern-Simons--Higgs theory in $2+1$ dimensions directly
generalising the Abelian model proposed in \cite{Hong:1990yh} and \cite{Jackiw:1990aw}. These solutions are
constructed numerically, and sit above the previously known~\cite{Hong:1990yh,Jackiw:1990aw} Abelian embeddings.
Our study is directed in three main directions:
\begin{itemize}
\item
To investigate the dependence of the energies of the non Abelian vortices on the various parameters characterising
them. Some of these parameters, denoted as $p_1\,,p_2$ and $\al$, are the asymptotic values of the various fields
describing the model, and since they are not independent of each other -- such relations being only seen via the
numerical process -- we have chosen to focus of the most convenient one, namely $p_2$, corresponding to the asymptotic
value of the electric potential. In particular, the known
Abelian embedding solutions 
possess the value $p_2=0$. Physically, the most significant parameter is the
Higgs selfinteraction strength $\nu$ of the symmetry breaking Higgs potential. We find it very convenient and
interesting therefore to compare the energies of the non Abelian vortices with those of the
Abelian ones {\it versus} the coupling $\nu$ in the $p_2=0$ case (see Fig.~8). We observe that for a range of values of
$\nu$ the energies of the non Abelian vortices are lower than those of their Abelian counterparts. But the latter being
topologically stable, this indicates that the corresponding non Abelian vortices are also stabilised by this bound.

The question of stability is a subtle one. Normally, when a new field is introduced in a classical system, the energy of
the system becomes smaller than the original one. Examples of this are the Skyrme model after
(diagonal) gauging with $SU(2)$~\cite{nonlinearity} or with
$U(1)$~\cite{Piette:1997ny,Radu:2005jp}, and the
$U(1)$ gauging of the Goldstone soliton on $\R^2$~\cite{Arthur:1998nh}. In both these examples the gauged soliton is
topologically stable against its own energy lower bound. On the other hand, when the system supporting a topologically
stable soliton of the (purely magnetically) $U(1)$ gauged soliton is augmented with an electric field such that the
topological lower bound on the energy of the new system is still bounded only by the lower bound of the purely magnetic
system, then it turns out that the energy of the electrically charged system is higher than the purely magnetic charged
one. This is the situation in the case at hand. The Abelian subsystem has a topological lower bound which remains the
only bound valid when the non Abelian degrees of freedom are introduced. One would therefore expect that the energy of
the non Abelian vortex should be higher than the Abelian embedding. Hence we find it significant that for a certain
range of the parameter $\nu$ the energy of the non Abelian vortex is lower than that of the Abelian one.

Apart from the question of stability illustrated by the branch structure of Figure~8, there is another interesting
property of the dependence of the energy on the parameters fixing the numerical solution, illustrated by the  branch
structure of Figure~9. The latter describes a branch of solutions that is disconnected from the Abelian branch, plotted
against the parameter $p_2$. This phenomenon is strictly one that appears for large values of $\nu$. While the
generic solutions are found to be connected to the Abelian embeddings, out of which they grow as their nonabelianness
manifests itself, in the high $\nu$ regime there appear disconnected branches. An explanation for this may be the fact
that as $\nu$ grows, the contribution of the Higgs potential term must vanish, resulting in the constraint $|\vec\f|^2=1$.
Thus the dynamics changes from that of a Higgs model to one of a $O(3)$ sigma model. As it happens, the $U(1)$ gauged
embedding of this sigma model does indeed support topologically stable solitons~\cite{Schroers:1995ns}, so that the
energy of the vortices of its $SO(3)$ gauged extensions are also bounded from below, albeit with higher
energy than the $SO(2)$ gauged soliton~\cite{Schroers:1995ns}. This passage from a Higgs model to a sigma model has
a noteworthy precedent, namely that of the sphalerons~\cite{Klinkhamer:1984di} of the standard model in the high
Higgs coupling regime, where the limiting solutions describe the bi-sphalerons~\cite{Brihaye:1993qv} coinciding with
the (right) $SU(2)$ gauged techniskyrmions~\cite{Eilam:1985tg} of the $O(4)$ sigma model.
In the  bi-sphaleron case, this is associated with the excitation of some extra-function in the ansatz of the matter
fields. Therefore, we anticipate the possible existence of new solutions of the model considered in this work as well,
which would be found beyond the truncation (\ref{trunc0}).

\item
Given that the only topological charge in this model is the magnetic charge, or the vortex number pertaining to the
Abelian subsystem, it is important to find some other global quantity that characterises the non Abelian vortices
exclusively. This is the angular momentum of the vortex, which in addition to the globally defined magnetic and electric
charges pertaining to the Abelian embedding and the energy, provides a global quantity that characterises the non
Abelian vortex. This quantity differs essentially in the Abelian and the non Abelian cases (see Eq. \re{J1}) and gives a
quantitative measure of the nonabelianness of the $SU(2)$ vortices.

The value of the angular momentum in the Abelian cases is independent of the coupling constant
$\nu$. In that case it is also independent of the energy, whether or not the value of $\nu$ is the critical
one when the Bogomol'nyi bound is saturated as in \cite{Hong:1990yh,Jackiw:1990aw}. In the non Abelian case by contrast,
the angular momentum does depend on the energy. This yields a mass {\it
versus} spin plot (see Figure~6). Not unexpectedly, it turns out that the only spinless solutions are those with vanishing
mass, {\it i.e.} trivial solutions with vanishing static energy. For a given value of $\nu$, namely for a given
physical model, this spin--energy behaviour is studied again, when varying our
favoured parameter $p_2$. We observe that for small values of $\nu$ the spin increases with energy, while for large $\nu$
there exist regions where the spin decresases with increasing energy. Concerning this, we recall our comment above
 where it is pointed out that the Higgs model at hand becomes a sigma model in the limit of very large $\nu$.

\item
Since the solutions we constructed pertain to the model in which the Higgs coupling constant $\nu$ is not restricted to
the critical Bogomol'nyi value, it is reasonable to inquire about the properties of the Abelian embedding vortices with
respect to $\nu$. It turns out that there exist both attractive and repulsive phases of like charged vortices for
non critical values of $\nu$ for $\nu<\nu_{critical}$ and $\nu>\nu_{critical}$ respectively, with non interacting
vortices for $\nu=\nu_{critical}$. This is also not surprising and is the same as the situation is for the usual Abelian
Higgs vortices.
\end{itemize}
Finally we comment on the reason for our choice of the simplest $SU(2)$ CS-H model, unlike the $SU(N)$ model of
\cite{deVega:1986hm} and the more general supersymmetry
inspired models employed in \cite{Aldrovandi:2007nb,Collie:2008za}. In our view, the present
$SO_{\pm}(4)\equiv SU_{\pm}(2)$ CS-H model in $2+1$ dimensions is the first member of a hierarchy of $SO_{\pm}(D+2)$
CS-H model in $D+1$ dimensions. It is planned to study the $D=4$ and $6$ examples in the near future.
Also, on general grounds, we expect the basic features of the model considered here to be generic for CS-H
configurations with nonabelian matter fields, and thus to give an idea of the situation in a more general case.

More immediately, we intend to revisit the problem of the present model augmented with the $SU(2)$ YM term,
to construct the corresponding non Abelian vortices numerically. Apart from its intrinsic value, such a
numerical investigation would reveal some detailed properties of the analytic results of \cite{Spruck:2008bm}.
The latter work was carried out in the
context of giving a rigorous proof of the result of Julia and Zee~\cite{Julia:1975ff} which states that the vortices of
the Abelian Higgs model in $2+1$ dimensions cannot carry electric charge. These authors went further to extend the
proof of the Julia and Zee theorem of the Abelian Higgs model, to the $SU(2)$ non Abelian Higgs model
in $2+1$ dimensions, {\it i.e.} that the electric charge vanishes also in that case. One would expect that in the limit
of the CS term vanishing the (non Abelian) electric field vanishes.

\medskip
\medskip

\noindent
{\bf\large Acknowledgements} This work is carried out in the framework of Science Foundation Ireland (SFI) project
RFP07-330PHY and 
under project FIS2006-12783-C03-02 of the Spanish Education and Science
Ministry. 
 The work of ER was supported by a fellowship from the Alexander von Humboldt Foundation.


\begin{small}

\end{small}

\end{document}